\documentclass[onecollarge,natbib]{svjour2}
\bibpunct{[}{]}{;}{n}{}{,} 
\smartqed  
\usepackage{graphicx}
\journalname{Few Body Systems}
\begin{document}

\title{Nucleon-to-pion transition distribution amplitudes: a challenge for \={P}ANDA 
}


\author{B.~Pire$^1$  \and  K.~Semenov-Tian-Shansky$^2$  \and L.~Szymanowski$^3$ }


\institute{    \at
$^1$ CPhT, {\'E}cole Polytechnique, CNRS, F-91128 Palaiseau, France. 
\at
$^2$  IFPA, d\'{e}partement AGO,  Universit\'{e} de  Li\`{e}ge, 4000 Li\`{e}ge,  Belgium.
\email{ksemenov@ulg.ac.be}
\at
$^3$  National Centre for Nuclear Research (NCBJ), Warsaw, Poland.
}

\date{}

\maketitle

\begin{abstract}
Baryon-to-meson Transition Distribution Amplitudes (TDAs) appear
as building blocks in the collinear factorized description of amplitudes
for a class of  hard exclusive reactions, prominent examples being hard exclusive pion
electroproduction off a nucleon in the backward region and baryon-antibaryon annihilation
into a pion and a lepton pair or a charmonium.
Baryon-to-meson TDAs extend both the concepts of generalized parton
distributions (GPDs) and baryon distribution amplitudes (DAs) encoding
 valuable complementary information on the hadronic structure.
We review the basic properties of baryon-to-meson TDAs and discuss the
perspectives for the experimental access with the \={P}ANDA detector.
\end{abstract}

\section{Introduction}
\label{Sec_introTDAs}

The \={P}ANDA experiment at GSI-FAIR
\cite{PANDAprogramm,Wiedner:2011mf}
is being built to address fundamental questions in hadronic physics. Although much progress
has been achieved in the domain of the deep structure of nucleons thanks
to lepton beam initiated reactions, the interior of hadrons is still a frontier
domain of the current understanding of QCD dynamics.
In this respect, accessing the transition distribution amplitudes
in specific exclusive reactions at \={P}ANDA is an important goal to progress in our understanding of
quark and gluon confinement.

The leading twist-$3$ baryon to meson (antibaryon to meson) TDAs are defined through
baryon (antibaryon)-meson matrix elements of
the nonlocal three quark (antiquark) operators on the light cone
\cite{Lepage:1980,Chernyak:1983ej,Stefanis_DrNauk}:
\begin{equation}
\hat{O}^{\alpha \beta \gamma}_{\rho \tau \chi}( \lambda_1 n,\, \lambda_2 n, \, \lambda_3 n)
 =
\varepsilon_{c_{1} c_{2} c_{3}}
\Psi^{c_1 \alpha}_\rho(\lambda_1 n)
\Psi^{c_2 \beta}_\tau(\lambda_2 n)
\Psi^{c_3 \gamma}_\chi (\lambda_3 n),
\label{operators}
\end{equation}
where
$\alpha$, $\beta$, $\gamma$
stand for quark (antiquark) flavor indices and
$\rho$, $\tau$, $\chi$
denote the Dirac spinor indices. Antisymmetrization stands over the color group indices
$c_{1,2,3}$.
Gauge links in
(\ref{operators}) are omitted by considering the light-like gauge
$A \cdot n=0$.
These non-perturbative objects first considered in
\cite{Frankfurt:1999fp},
share common features both with baryon DAs (that are defined as the baryon-to-vacuum matrix elements of the same operator
(\ref{operators}))
and with GPDs since the matrix element in question depends on the longitudinal momentum transfer between a baryon and a meson
characterized by the skewness variable
$\xi$.

Baryon-to-meson TDAs arise in the collinear factorized description of several
hard exclusive reactions such as backward electroproduction of mesons off nucleons
\cite{Lansberg:2007ec,Lansberg:2011aa}
that can be studied at JLab
\cite{Kubarovskiy:2012yx}
and COMPASS.
Future  \={P}ANDA  facility at GSI-FAIR makes it possible to access different classes of
reactions that may be described in terms of baryon-to-meson TDAs, such as
$N \bar{N}$
annihilation into a lepton pair in association with a light meson
\cite{Lansberg:2007se,Lansberg:2012ha}
 or into a heavy quarkonium and a meson
\cite{Pire:2013jva}.
This will increase the experimental support profiting from the outstanding exclusive detection capabilities
of \={P}ANDA and allow to check the universality of baryon-to-meson TDAs combining information on the space-like regime
(from JLab) and the time-like regime (from  \={P}ANDA).

Although baryon to meson TDAs can be introduced for all types of baryons and mesons,
we mostly consider here the simplest case of nucleon-to-pion
($\pi N$)
TDAs. Below we  summarize the fundamental requirements for
$\pi N$
TDAs which follow from the symmetries of QCD established in Refs.
\cite{Pire:2010if,Pire:2011xv,Lansberg:2011aa}.

\begin{itemize}
\item  For a given flavor contents, the spin decomposition of the leading twist-$3$
$\pi N$
TDA involves eight real valued invariant functions
$V_{1,2}^{\pi N}$, $A_{1,2}^{\pi N}$, $T_{1,2,3,4}^{\pi N}$,
each depending on the  longitudinal momentum fractions
$x_i$ ($\sum_{i=1}^3 x_i=2\xi$),
skewness parameter
$\xi$
and the momentum transfer squared
$\Delta^2 $,
as well as on the factorization scale $\mu^2$.

\item The support of $\pi N$ TDAs in three longitudinal momentum fractions
$x_i$
is given by the intersection of  the stripes
$-1+\xi \le x_i \le 1+\xi$ ($\sum_{i=1}^3 x_i=2\xi$).
One can distinguish the Efremov-Radyushkin-Brodsky-Lepage-like (ERBL-like) domain and two types of
Dokshitzer-Gribov-Lipatov-Altarelli-Parisi-like (DGLAP-like) domains.

\item The evolution properties of
$\pi N$
TDAs are described by the appropriate generalization
\cite{Pire:2005ax}
of the  ERBL/ DGLAP evolution equations specific for different domains in
$x_i$.

\item Similarly to the GPD case, the underlying Lorentz symmetry results in the polynomiality
property for the Mellin moments of
$\pi N$
TDAs in the longitudinal momentum fractions
$x_i$.
To ensure the polynomiality and the restricted
support properties for
$\pi N$
TDAs one can employ the spectral representation in terms of quadruple distributions
\cite{Pire:2010if},
which generalizes for the TDA case Radyushkin's double distribution representation
for GPDs.

\item  Contrary to GPDs,
$\pi N$
TDAs lack a comprehensible forward limit
($\xi=0$).
It is illuminating to consider the alternative limit
$\xi=1$
in which
$\pi N$ TDAs
are constrained by chiral dynamics and crossing due to the soft pion theorem.

\item In order to satisfy the polynomiality condition in its complete form, the spectral
representation for
$\pi N$
TDAs should be supplemented with a $D$ term-like contribution.
The simplest possible model for such a
$D$
term is the contribution of the cross-channel nucleon exchange computed in
\cite{Pire:2011xv}.
\end{itemize}

\section{Pion production in association with a high invariant mass lepton pair in $\bar{p}N$ annihilation}
One of the important aims of the experimental program of \={P}ANDA
will be the study of the nucleon electromagnetic form factor in the time-like region in
nucleon-antinucleon annihilation into a lepton pair. Outside the resonance region (for
high invariant mass
$q^2 \equiv Q^2$
of the lepton pair) the description of the nucleon electromagnetic form factors can be provided
by the methods of pertubative QCD
\cite{Chernyak:1983ej}.

In
\cite{Lansberg:2007se,Lansberg:2011aa}
we argue that a tempting possibility to apply similar pQCD methods for exclusive reaction
is to consider also  the nucleon-antinucleon  annihilation into a high invariant mass lepton pair
in association with a light meson
${\mathcal{M}}=\{\pi,\,\eta,\, \rho, \omega,\, ... \}$:
\begin{equation}
\bar{N} (p_{\bar{N}},s_{\bar{N}})+  N (p_N,s_N) \rightarrow \gamma^*(q)+ {\mathcal{M}}(p_{\mathcal{M}}) \rightarrow \ell^+(p_{\ell^+}) +
\ell^-(p_{\ell^-}) + {\mathcal{M}}(p_{\mathcal{M}}).
\label{BarNNannihilation reaction}
\end{equation}

\begin{figure*}
\centering
  \includegraphics[width=0.30\textwidth]{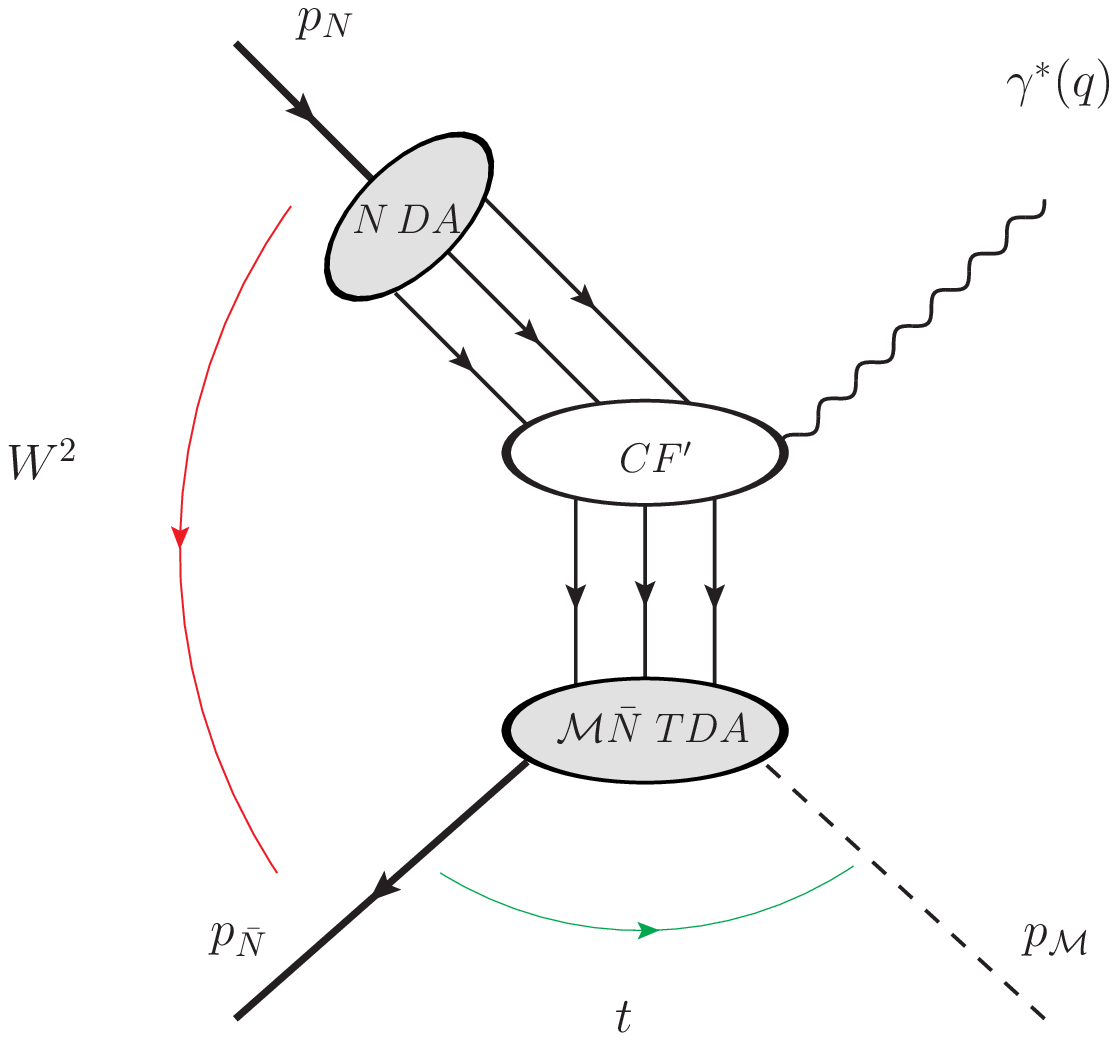}
   \includegraphics[width=0.30\textwidth]{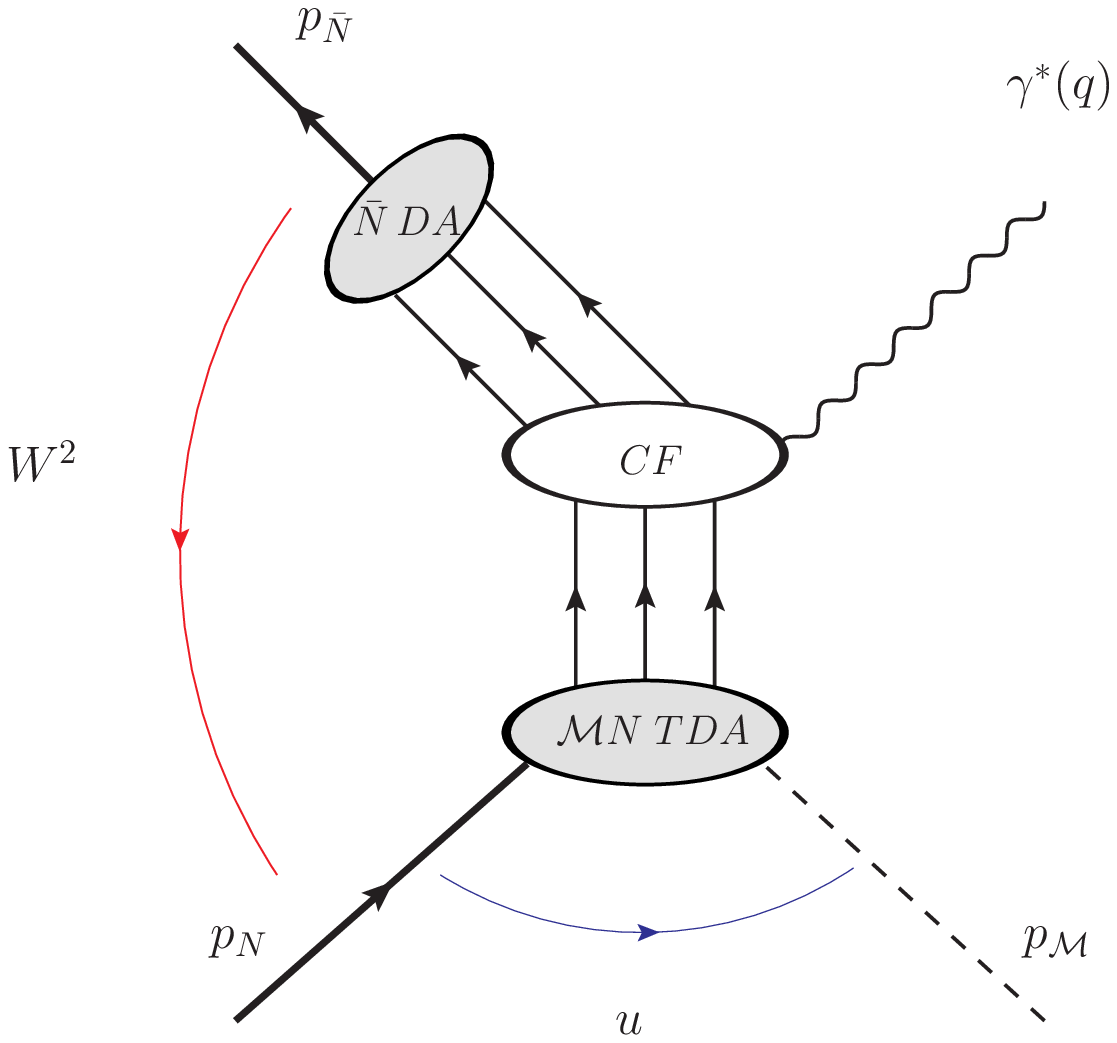}
\caption{Two possibilities for  collinear factorization  of the annihilation
     process
$N \bar{N} \to \gamma^*(q) \mathcal{M}(p_\mathcal{M})$.
{\bf Left panel:}  forward  kinematics
($t \sim 0$) .
{\bf Right panel:}
backward  kinematics
($u \sim 0$).
$\bar{N}(N)$
DA stands for the distribution amplitude
of antinucleon (nucleon);
$\mathcal{M} N (\mathcal{M} \bar{N})$
TDA stands for the transition distribution amplitude from a nucleon (antinucleon) to a meson;
CF and CF' denote hard subprocess amplitudes (coefficient functions). }
\label{Fig_DileptonPion_BkwFwd}
\end{figure*}

The factorization mechanism for
(\ref{BarNNannihilation reaction})
suggested in
\cite{Lansberg:2007se}
is shown on Fig.~\ref{Fig_DileptonPion_BkwFwd}.
We choose the
$z$
axis along the colliding
$N \bar{N}$
with positive direction along the antinucleon beam. We introduce the
$t$-
and
$u$-channel light-cone vectors
$n^t$, $p^t$;  $n^u$, $p^u$
and define the
$t$
and
$u$
channel skewness variables
$\xi^t \equiv - \frac{(p_{\mathcal{M}}-p_{\bar{N}}) \cdot n^t}{(p_{\mathcal{M}}+p_{\bar{N}}) \cdot n^t} $,
$\xi^u \equiv - \frac{(p_{\mathcal{M}}-p_N) \cdot n^u}{(p_{\mathcal{M}}+p_N) \cdot n^u} $.
The amplitude of the
$N \bar{N} \to \gamma^* \mathcal{M}$
subprocess is presented as a convolution of the hard part computed by means of perturbative QCD with nucleon
DAs and nucleon to pion TDAs encoding the soft dynamics. The factorization is supposed to be achieved in
two distinct kinematical regimes:
\begin{itemize}
\item  the near forward regime
($s=(p_N+ p_{\bar{N}})^2 \equiv W^2$, $Q^2$
large with
$\xi^t$
fixed; and
$|t|=|(p_{\mathcal{M}}-p_{\bar{N}})^2| \sim 0$)
this corresponds to the produced pion moving nearly in the direction of initial
$\bar{N}$ in  $\bar{N}N$
center-of-mass system(CMS). \item  the near backward regime
($s=(p_N+ p_{\bar{N}})^2\equiv W^2$, $Q^2$
large with $\xi^u$
fixed;
$|u|=|( p_{\mathcal{M}}-p_N)^2| \sim 0$)
this corresponds to the produced pion moving nearly in the direction of initial
$N$
in
$\bar{N}N$
CMS.
\end{itemize}
The suggested reaction mechanism should manifest itself through the distinctive forward
and backward peaks of the
$N \bar{N} \to \gamma^* \mathcal{M}$
cross section. The charge conjugation invariance results in perfect symmetry between
the two kinematical regimes.

For definiteness, we below consider the case of the meson being a pion. Within the
factorized approach of
\cite{Lansberg:2007se},
at leading order in
$\alpha_s$,
the amplitude of
$N \bar{N} \to \gamma^* \pi$
$\mathcal{M}_{\lambda}^{s_N s_{\bar{N}}}$
reads
\begin{equation}
\mathcal{M}_{\lambda}^{s_N s_{\bar{N}}}=
\mathcal{C} \frac{1}{Q^4}
\Big[
\mathcal{S}_{\lambda}^{s_N s_{\bar{N}}}
\mathcal{I}(\xi, \Delta^2)-
\mathcal{S'}_{\lambda}^{s_N s_{\bar{N}}}
\mathcal{I}'(\xi, \Delta^2)
\Big],
\label{Def_ampl_M}
\end{equation}
where
$
{ \cal C}=-i \frac{(4 \pi \alpha_s)^2 \sqrt{4 \pi \alpha_{em}} f_N^2}{54 f_\pi} ;
$
with
$f_N \approx 5.0 \cdot 10^{-3}$ GeV$^2$
standing for the nucleon wave function normalization constant,
$f_\pi=93$ MeV -- pion weak decay constant and $\alpha_s$ is set to $\alpha_s=\bar{\alpha}_s \equiv 0.3$.
The spin structures
(\ref{Def_ampl_M})
are defined as
$ \mathcal{S}_{\lambda}^{s_N s_{\bar{N}}} \equiv
\bar{V}(p_{\bar{N}},s_{\bar{N}}) \hat{\epsilon}^*(\lambda) \gamma_5 U(p_N,s_N)
$;
$\mathcal{S'}_{\lambda}^{s_N s_{\bar{N}}} \equiv
\frac{1}{M}
\bar{V}(p_{\bar{N}},s_{\bar{N}}) \hat{\epsilon}^*(\lambda) \hat{\Delta}_T \gamma_5 U(p_N,s_N),
$
where $V$
and
$U$
are the usual nucleon Dirac spinors and the Dirac ``hat'' notation
$\hat{v}= \gamma_\mu v^\mu$
is employed.
$\epsilon(\lambda)$
stands for the polarization vector of the virtual photon.
$\mathcal{I}$
and
$\mathcal{I'}$
denote the convolution integrals of
$\pi N$
TDAs and antinucleon DAs with the hard scattering kernels computed from the set of
$21$
relevant scattering diagrams
\cite{Lansberg:2007ec}.
The hard scattering kernels for backward
$p \bar{p} \to \gamma^* \pi$
differ from those for
$p   \gamma^*  \to \pi p$
by the replacement
$- i \varepsilon \to i \varepsilon$ in the corresponding
denominators. The averaged-squared amplitude for the process
(\ref{BarNNannihilation reaction})
then reads
\begin{eqnarray}
|\overline{\mathcal{M}^{ N \bar{N} \rightarrow \ell^+ \ell^- \pi }}|^2=
\frac{1}{4}
\sum_{s_N, \, s_{\bar{N}}, \, \lambda, \, \lambda'}
\mathcal{M}_{\lambda}^{s_N s_{\bar{N}}}
\frac{1}{Q^2}
e^2 {\rm Tr}
\left\{
\hat{p}_{\ell^-} \hat{\epsilon}(\lambda) \hat{p}_{\ell^+} \hat{\epsilon}^*(\lambda')
\right\}
\frac{1}{Q^2}
\left( \mathcal{M}_{\lambda'}^{s_N s_{\bar{N}}} \right)^*.
\label{WF_Cross_sec}
\end{eqnarray}
The differential cross section of the reaction
(\ref{BarNNannihilation reaction})
reads
\begin{eqnarray}
\frac{d \sigma}{du d Q^2 d \cos \theta_{\ell}}=
 \frac{\int d \varphi_\ell |\overline{\mathcal{M}^{N \bar{N} \rightarrow \ell^+ \ell^- \pi}}|^2 }
 {64 W^2 (W^2-4M^2) (2 \pi)^4},
 \label{Cross_sec_PANDA}
\end{eqnarray}
where
$\theta_{\ell}$
and
$\varphi_\ell$
are the lepton polar and azimuthal angles in
$\ell^+ \ell^- $
CMS.
To the leading twist accuracy, only the transverse polarization of the virtual photon
is contributing. Computing the relevant traces and integrating over the lepton  azimuthal  angle
one gets
\begin{eqnarray}
 \int d \varphi_{\ell} \, |\overline{\mathcal{M}^{ N \bar{N} \rightarrow \ell^+ \ell^- \pi }}|^2
\Big|_{\rm Leading \, twist}=
 2 \pi e^2(1+\cos^2 \theta_\ell)  \frac{1}{4} |\mathcal{C}|^2 \frac{2(1+\xi)}{\xi Q^8}
 \big( |\mathcal{I}|^2- \frac{\Delta_T^2}{M^2} |\mathcal{I}'|^2 \big).
\end{eqnarray}
The specific behavior in
$\cos^2 \theta_\ell$
of the cross section
(\ref{Cross_sec_PANDA})
together with the characteristic scaling behavior in
$1/Q$
may be seen as the distinctive features of the proposed factorization mechanism.

\label{Sec_Dilepton}

\begin{figure*}
\centering
  \includegraphics[width=0.41\textwidth]{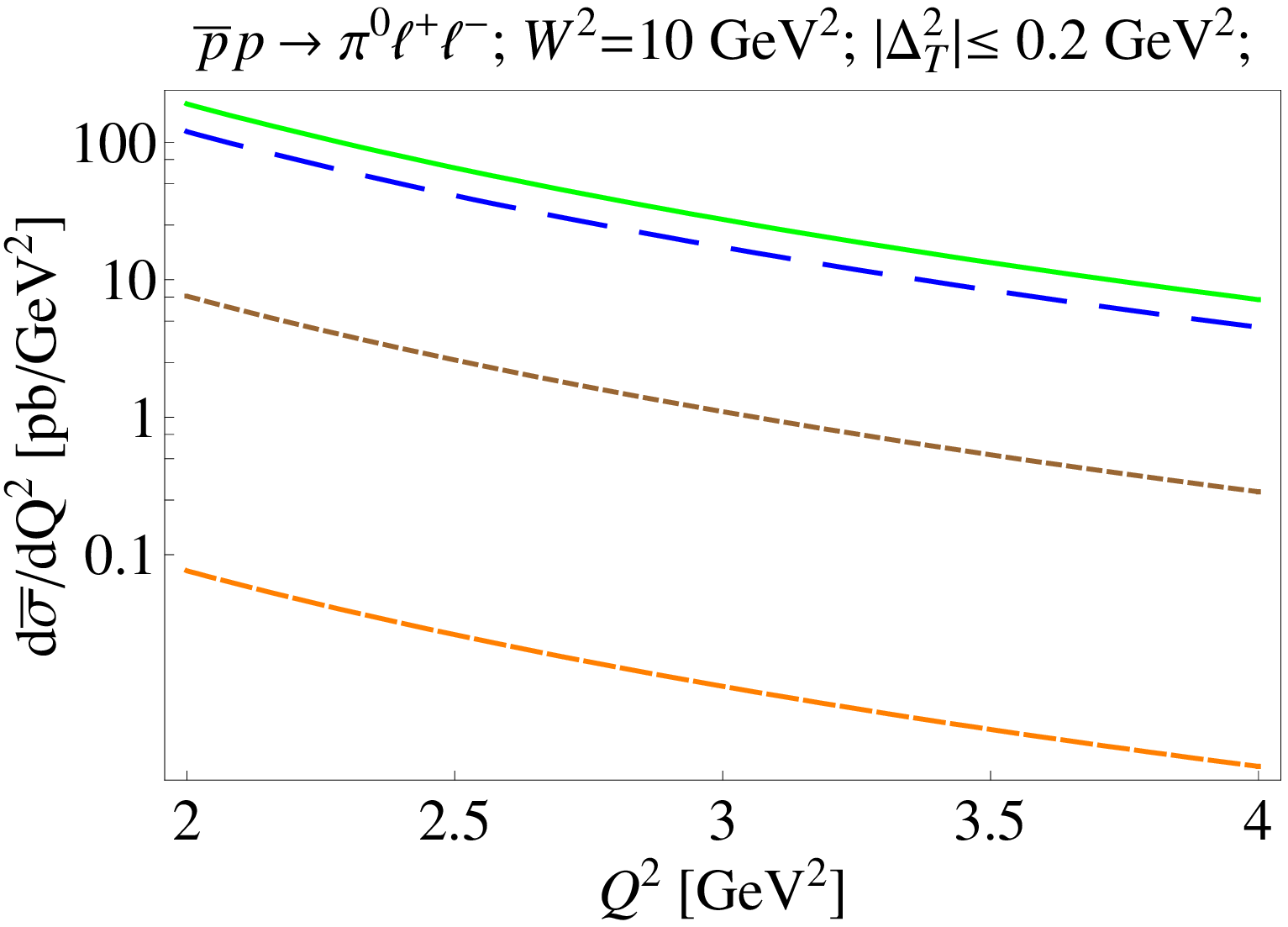}
   \includegraphics[width=0.41\textwidth]{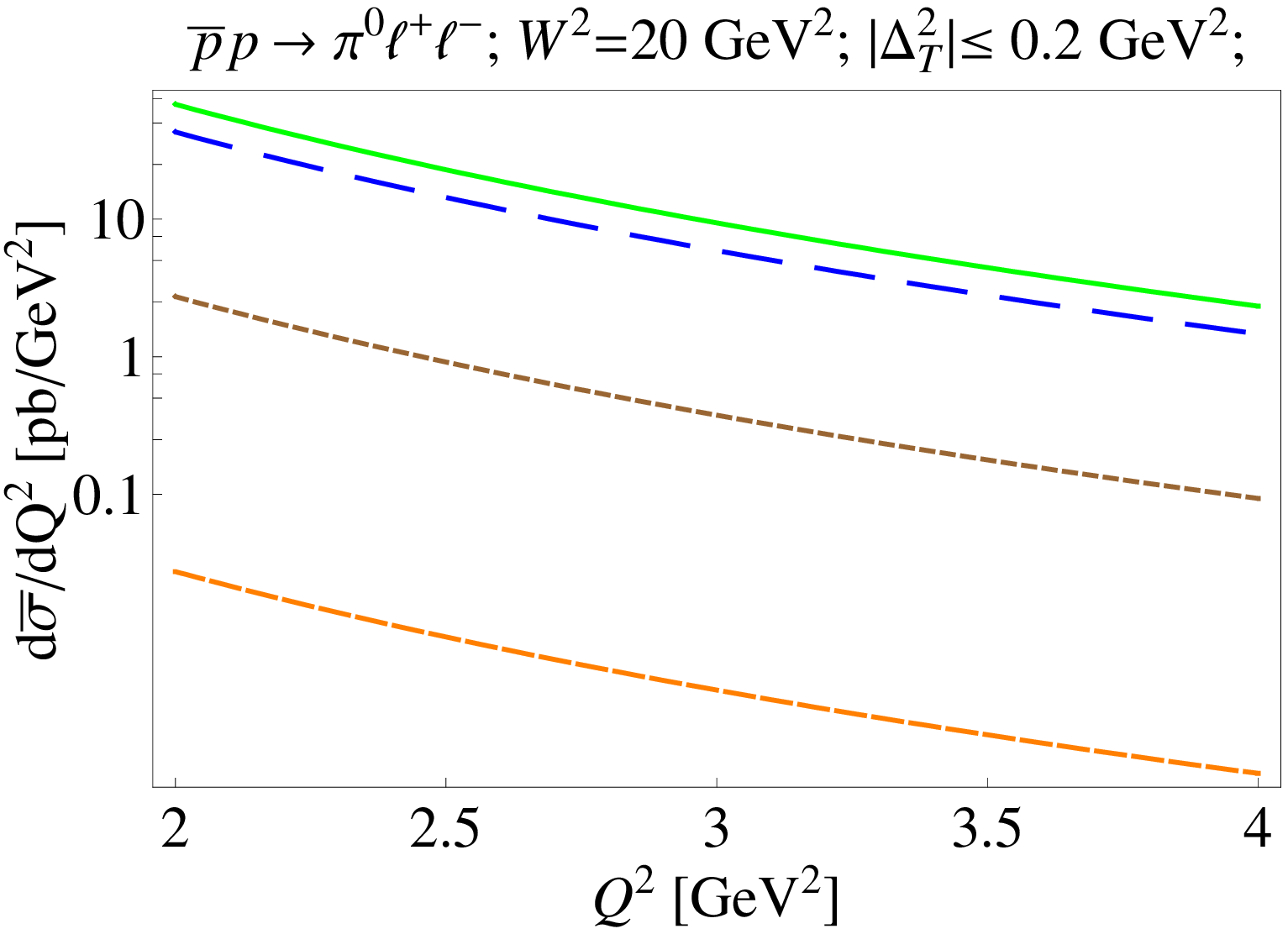}
\caption{ Integrated cross section
$d \bar{\sigma}  /dQ^2$
for
$\bar{p}p \rightarrow \ell^+\ell^- \pi^0$
as a function of
$Q^2$
for
$W^2=10$ GeV$^2$ and $W^2=20$ GeV$^2$
for various phenomenological nucleon DA solutions: COZ (long blue dashes);
KS (solid green line);   BLW NLO
(medium orange dashes) and NNLO modification~\cite{Lenz:2009ar} of BLW (short brown dashes).}
\label{Fig_Dilepton_BkwFwd_CS}       
\end{figure*}

On Fig.~\ref{Fig_Dilepton_BkwFwd_CS}
we present our estimates for the integrated cross section
\begin{equation}
\frac{d \bar{\sigma} }{dQ^2}({\Delta_T^2}_{\min}) \equiv \int_{u_{\min}}^{u_{\max}} du \int d \theta_{\ell} \frac{d \sigma}{du dQ^2 d \cos \theta_\ell }
\label{Def_itegrated_CS}
\end{equation}
for backward
$\bar{p}p \rightarrow \ell^+\ell^- \pi^0$,
as a function of
$Q^2$
for
$W^2=10$ GeV$^2$
and
$W^2=20$ GeV$^2$
within the nucleon pole exchange model
\cite{Pire:2011xv}
for
$\pi N$
TDAs. We have integrated the cross section
(\ref{Def_itegrated_CS})
over the bin in
$u$
in order to implement the effect of the cut
${\Delta_T^2}_{\min} \le \Delta_T^2 \le 0$
intended to focus on the backward kinematics regime. As the numerical input for the TDA model  we use different
phenomenological solutions for nucleon DAs: Chernyak-Ogloblin-Zhitnitsky (COS)
\cite{Chernyak:1987nv}
King-Sachrajda (KS)
\cite{King:1986wi},
Braun-Lenz-Wittmann (BLW) NLO
\cite{Braun:2006hz}
and NNLO modification~\cite{Lenz:2009ar}
of BLW. For most of the choices, the magnitudes of the cross sections are sufficient to be
measurable, with the luminosity foreseen at the \={P}ANDA experiment at GSI-FAIR.

\section{$J/\psi$ plus pion production in $\bar{p} N$ annihilation}
\label{Sec_Charmonium}

A major goal of the \={P}ANDA experiment is the study of the spectrum of charmonium states. Below we
argue that the concept of TDAs might be useful for understanding the non resonant background for
charmonium production.

\begin{figure*}
\centering
  \includegraphics[width=0.35\textwidth]{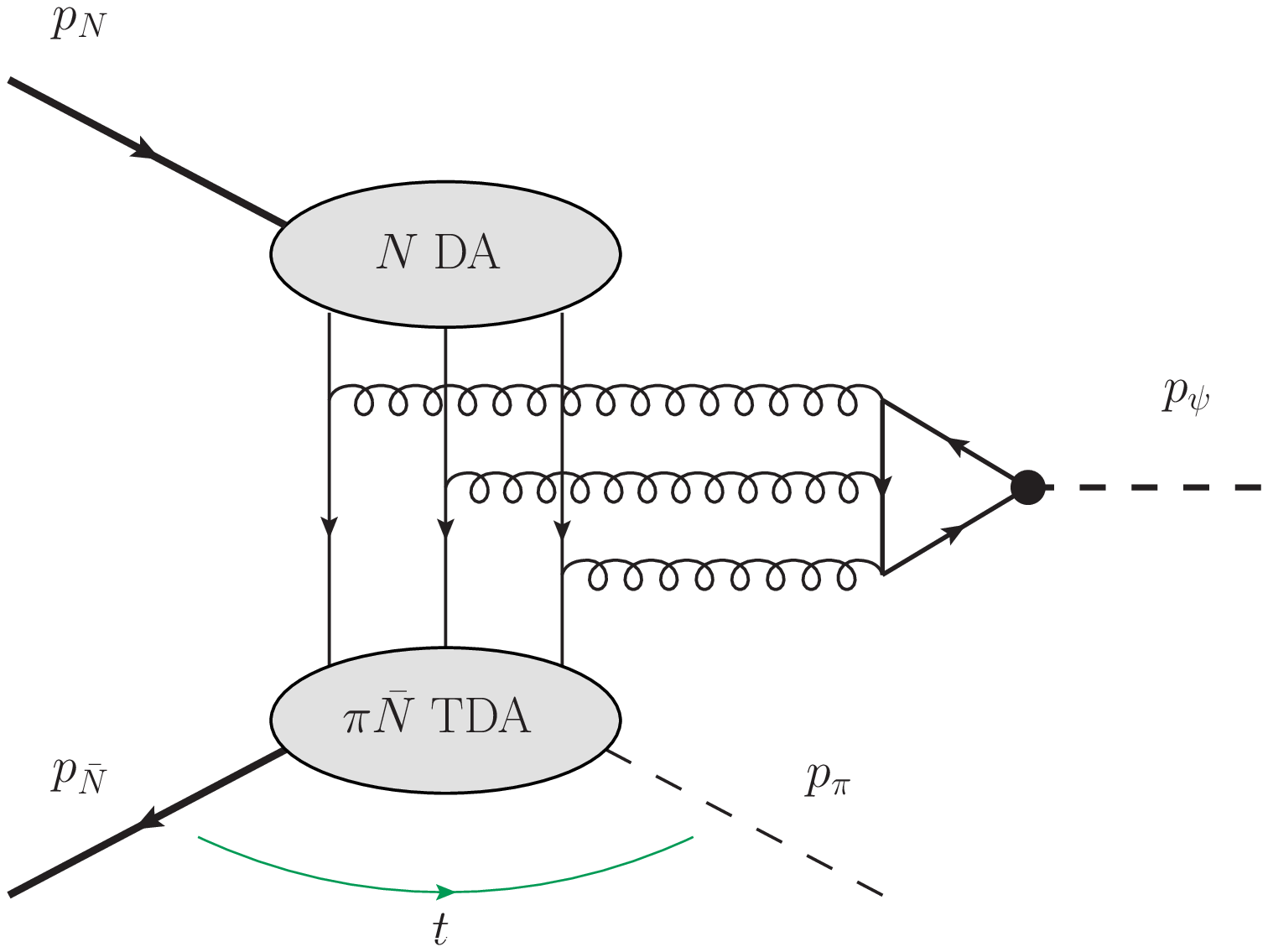}
   \includegraphics[width=0.35\textwidth]{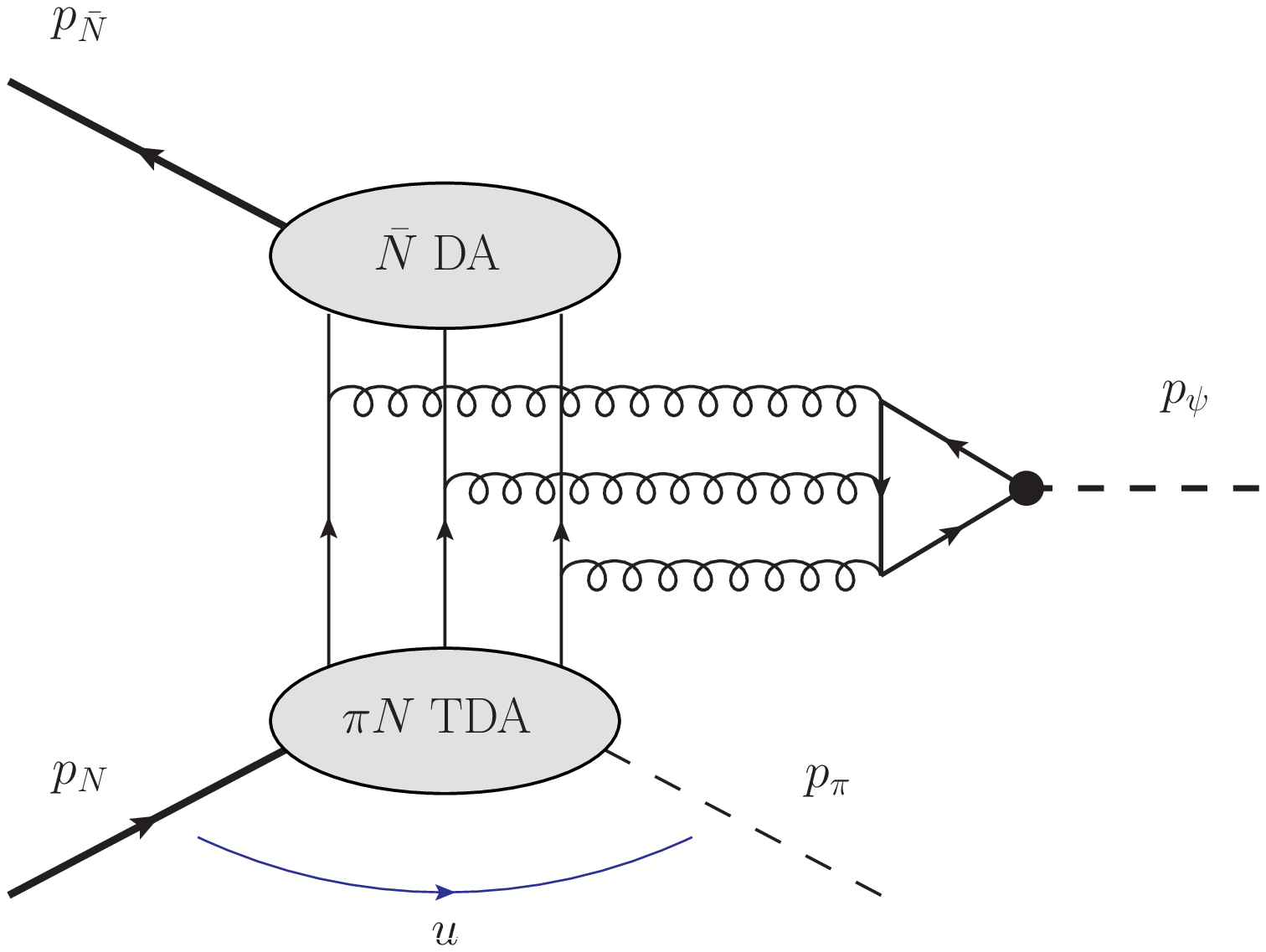}
\caption{Collinear factorization of the annihilation process
(\ref{reac}).
{\bf Left panel:} forward kinematics
($t \sim 0$).
{\bf Right panel:} backward kinematics
($u \sim 0$).
$\bar{N}(N)$
DA stands for the distribution amplitude of antinucleon (nucleon);
$\pi N (\pi \bar{N})$
TDA stands for the transition distribution amplitude from a nucleon (antinucleon) to a pion.}
\label{Fig_Charmonium_BkwFwd}       
\end{figure*}

The study of charmonium exclusive decays into hadrons historically has been one
of the first successful applications of perturbative QCD methods for hard
exclusive reactions.  The annihilation of the
$c \bar{c}$
pair into the minimal possible number of gluons producing quark-antiquark pairs which then
form outgoing hadrons was recognized to be the dominant mechanism. In
\cite{Pire:2013jva}
we extend the same framework for the description of nucleon-antinucleon annihilation into
the heavy quarkonia together with a pion:
\begin{equation}
  N (p_N) \;+ \bar N (p_{\bar N}) \; \to  J/\psi(p_{\psi})\;+\; \pi(p_{\pi}).
  \label{reac}
\end{equation}
The
$\bar{N} N$
center-of-mass energy squared
$s=(p_N+p_{\bar{N}})^2=W^2$
and the charmonium mass squared
$M_\psi^2$
introduce the natural hard scale for the problem. Similarly to the  nucleon-antinucleon annihilation into a
lepton pair and a light meson, it is assumed that the reaction
(\ref{reac})
admits a factorized description in the near forward
($t \equiv (p_\pi - p_{\bar{N}})^2 \sim 0$)
and  near backward
($u \equiv (p_\pi - p_N)^2 \sim 0$)
kinematical regimes. The corresponding mechanisms are presented on
Fig.~\ref{Fig_Charmonium_BkwFwd}.
Similarly to the case of
$N \bar{N}$
annihilation into a high invariant mass lepton pair in association with a pion the
$C$
invariance results in perfect symmetry between the forward and backward regimes of the reaction
(\ref{reac}).
Below we consider the backward regime. The
$z$
axis is chosen along the colliding
$N \bar{N}$
with positive direction along the antinucleon beam. We introduce the light-cone vectors satisfying
$2 p^u \cdot n^u=1$
and define the
$u$-channel skewness variable
$ \xi \equiv - \frac{(p_\pi-p_N) \cdot n^u}{(p_\pi+p_N) \cdot n^u} \simeq \frac{M^2_\psi}{2 W^2 -M^2_\psi}$.
Following
\cite{Chernyak:1987nv},
in our calculation we set the relevant masses to the average value
$ M_\psi  \simeq  2m_c   \simeq \bar{M} =3 \; {\rm GeV }$.
The physical kinematical domain for the reaction
(\ref{reac})
in the backward regime is determined by the requirement
${\Delta_T^2} \le 0$,
where
$
{\Delta_T^2}= \frac{1-\xi}{1+\xi} \left(\Delta^2- 2\xi \left[\frac{m_N^2}{1+\xi} -\frac{m_\pi^2}{1-\xi} \right] \right).
$

The leading order amplitude of the reaction
(\ref{reac})
from the mechanism presented on
Fig.~\ref{Fig_Charmonium_BkwFwd}
was computed in
\cite{Pire:2013jva}. It reads:
\begin{eqnarray}
&&
{\cal M}^{s_N s_{\bar{N}}}_\lambda
= {\cal C} \frac{1}{{\bar M}^5 } \Big[
\bar{V}(p_{\bar{N}},s_{\bar{N}} )\hat{\cal E}^*(\lambda) \gamma_5 U(p_N, s_N) {\cal J}(\xi,\Delta^2) \nonumber \\ &&
-\frac{1}{m_N}  \bar{V}(p_{\bar{N}},s_{\bar{N}} )\hat{\cal E}^*(\lambda) \hat{\Delta}_T \gamma_5 U(p_N, s_N) {\cal J}'(\xi,\Delta^2)
\Big].
\label{Amplitude_master}
\end{eqnarray}
Here
${\cal J}, \,{\cal J}'(\xi,\Delta^2)$
stand for the convolutions of the hard kernels with
$\pi N$
TDAs and antinucleon DAs.
Here
${\cal C}=(4 \pi \alpha_s)^3 \frac{f_N^2 f_\psi}{f_\pi}  \, \frac{10}{81}$,
where
$f_\psi=413 \pm 8$ MeV
is the normalization constant of the heavy quarkonia wave function. Its value is fixed from the charmonium leptonic
decay width
$\Gamma(J/ \psi \to e^+ e^-)$.
Within the suggested reaction mechanism it is the transverse polarization of charmonium that
is relevant to the leading twist accuracy. Summing over the transverse polarization of charmonium
and averaging over spins of initial nucleons we get
\begin{eqnarray}
|\overline{\mathcal{M}_{T}}|^2 \equiv \sum_{\lambda_T}  \frac{1}{4}
\sum_{s_N s_{\bar{N}}} {\cal M}^{s_N s_{\bar{N}}}_\lambda \left( {\cal M}^{s_N s_{\bar{N}}}_{\lambda'} \right)^*=
\frac{1}{4} |\mathcal{C}|^2 \frac{2(1+\xi)}{\xi {\bar{M}}^8} 
  \left( |\mathcal{J}(\xi, \Delta^2)|^2 - \frac{\Delta_T^2}{m_N^2} |\mathcal{J}'(\xi, \Delta^2)|^2 \right).
\end{eqnarray}

The leading twist differential cross section of
$N + \bar{N} \to J/\psi + \pi$
then reads
\begin{eqnarray}
\frac{d \sigma}{d \Delta^2}= \frac{1}{16 \pi \Lambda^2(s,m_N^2,m_N^2) } |\overline{\mathcal{M}_{T}}|^2,
\label{CS_def_delta2}
\end{eqnarray}
where
$\Lambda(x,y,z)= \sqrt{x^2+y^2+z^2-2xy-2xz-2yz}$.

\begin{figure*}
\centering
  \includegraphics[width=0.40\textwidth]{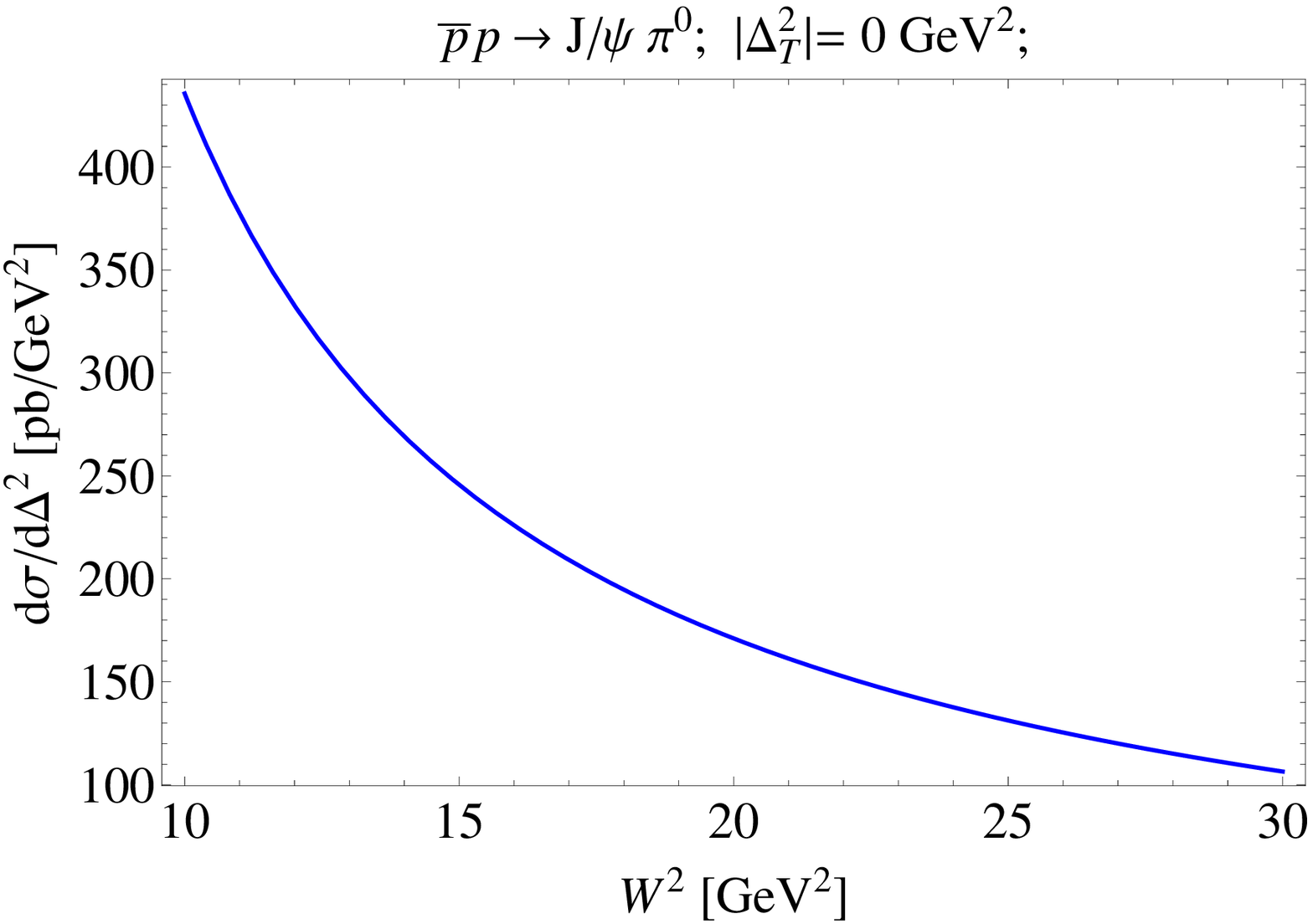}
   \includegraphics[width=0.40\textwidth]{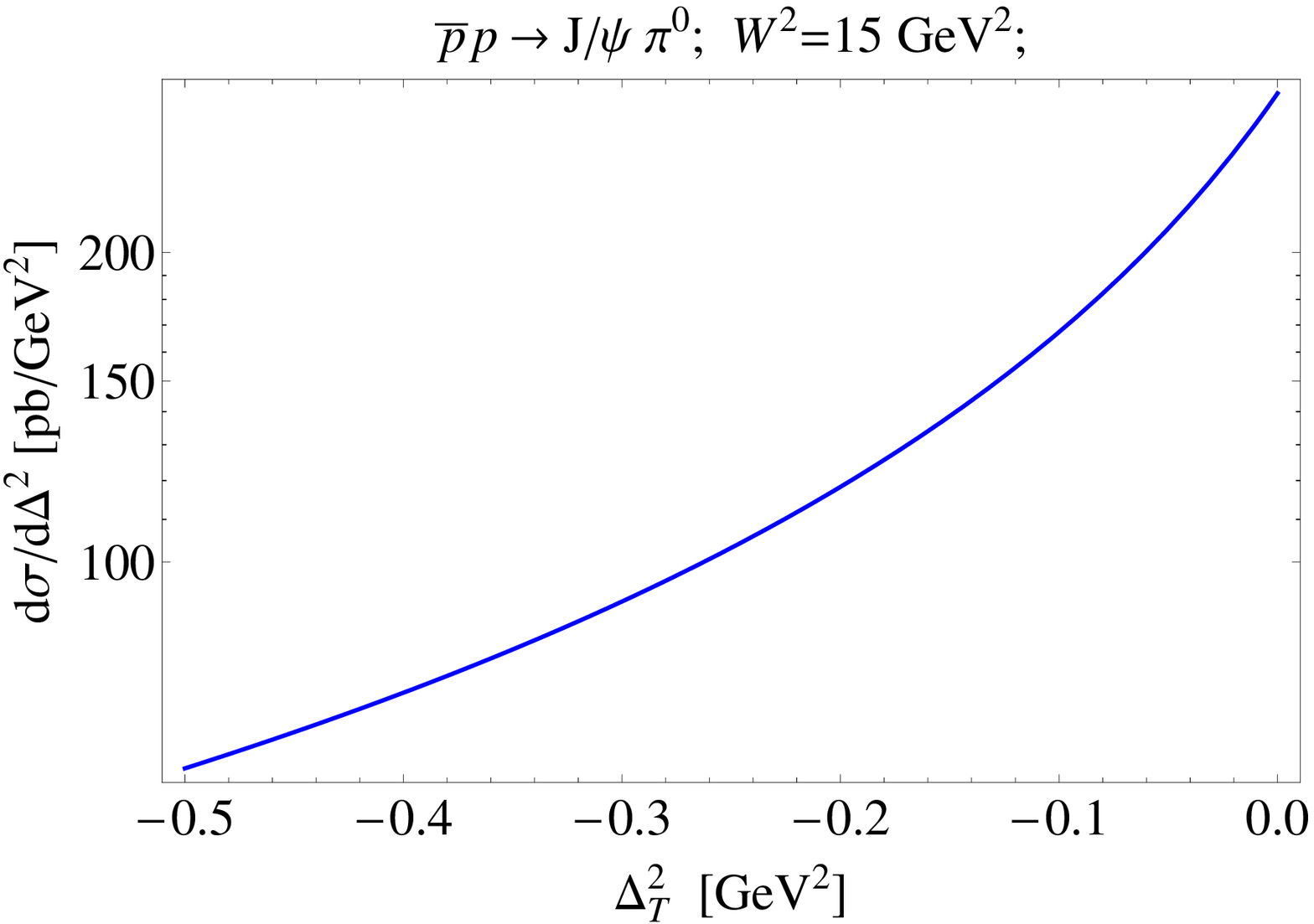}
\caption{{\bf Left panel:} Differential cross section
$\frac{d \sigma}{d \Delta^2}$ for $p \bar{p} \to J/\psi \,  \pi^0$
as a function of
$W^2$
for
$\Delta_T^2=0$. {\bf Right panel:} Differential cross section
$\frac{d \sigma}{d \Delta^2}$ for $p \bar{p} \to J/\psi \,  \pi^0$
as a function of
$\Delta_T^2$
for
$W^2=15$ GeV$^2$.}
\label{Fig_Charmonium_BkwFwd_CS}       
\end{figure*}

On Fig.~\ref{Fig_Charmonium_BkwFwd_CS}
we show our predictions for the differential cross section of backward
$p \bar{p} \to J/\psi \,  \pi^0$
as a function of
$W^2$
for
$\Delta_T^2=0$
and as a function of
$\Delta_T^2$
for fixed value of CMS energy squared
$W^2=15$ GeV$^2$
within the simple  nucleon pole exchange model  for
$\pi N$ TDAs
\cite{Pire:2011xv}.
As the numerical input for the model we use BLW NLO phenomenological
solution for nucleon DAs
\cite{Braun:2006hz}.
Note that the decay width shows strong dependence on
$\alpha_s: \, \sim \alpha_s^6$.
There is no unique opinion in the literature on the value of the strong coupling for the
gluon virtualities in question. To get a rough estimate of the cross section we fix the
strong coupling for a given phenomenological solution in a way that it reproduces the experimental
value for the charmonium decay width into proton-antiproton
$\Gamma(J/\psi \to p \bar{p})$.
In particular case of the  BLW NLO solution we then have to take
$\alpha_s=0.44$.
The obtained values of cross sections give hope of experimental accessibility of the reaction with \={P}ANDA.
Also our predictions are consistent with the recent estimates of
\cite{Lin:2012ru}
obtained within a fully non-perturbative effective hadronic theory.

\section{Conclusions}
\label{Sec_Conclusions_TDA}

Baryon-to-meson TDAs are new non-perturbative objects which have been designed to help us scrutinize the inner structure
of nucleons. Experimentally, one may access  TDAs both in the space-like domain
with backward electroproduction of mesons at JLab and COMPASS
and in the time-like domain in antiproton nucleon annihilation processes.
Extracting TDAs from space-like and time-like reactions will be a stringent test of their universality
\cite{Muller:2012yq}, and hence of the
factorization property of hard exclusive amplitudes. This hopefully will help us to disentangle the complex dynamics of quark and gluon confinement
in hadrons.

\begin{acknowledgements}

We acknowledge useful discussions with T. Hennino, J.P. Lansberg, B. Ma, F. Maas, B Ramstein and M. Zambrana.
This work is supported in part by the Polish Grant NCN
No DEC-2011/01/D/ST2/03915,  the Joint Research Activity "Study of Strongly
Interacting Matter" (acronym HadronPhysics3, Grant 283286) under the Seventh
Framework Programme of the European Community and by the COPIN-IN2P3 Agreement.
\end{acknowledgements}


\end{document}